\documentclass[graybox, envcountchap]{svmult}

\usepackage{mathptmx}        
\usepackage{amsmath}
\usepackage{amssymb}
\usepackage{color}
\usepackage{helvet}          
\usepackage{courier}         
\usepackage{dirtree}

\usepackage{makeidx}        
\usepackage{graphicx}        
\usepackage{subfig}
\usepackage{ulem}

\usepackage{multicol}        
\usepackage[bottom]{footmisc}

\usepackage{hyperref}        
\hypersetup{colorlinks=true,urlcolor=blue,citecolor=blue,linkcolor=blue}

\usepackage[misc]{ifsym}
\usepackage{cite}

\makeindex             

\definecolor{valecol}{rgb}{0,0.5, 1.}

\def\d{{\rm d}}

\graphicspath{{./figs/}}

\begin{document}


\title{The tension in the absolute magnitude of Type Ia supernovae}
\author{David Camarena and Valerio Marra}
\institute{David Camarena (\Letter) \at Department of Physics and Astronomy, University of New Mexico, Albuquerque, NM 87106, USA, \email{dacato115@gmail.com}
\and Valerio Marra \at Núcleo de Astrofísica e Cosmologia \& Departamento de Física, Universidade Federal do Espírito Santo, 29075-910, Vitória, ES, Brazil, \email{valerio.marra@me.com}}
%
%
\maketitle

\abstract{
This study aims to elucidate the tension in the Hubble constant ($H_0$), a key metric in cosmology representing the universe's expansion rate. Conflicting results from independent measurements such as the Planck satellite mission and the SH0ES collaboration have sparked interest in exploring alternative cosmological models. We extend the analysis by SH0ES to an arbitrary cosmographic model, obtaining a competitive local $H_0$ determination which only assumes the standard flat $\Lambda$CDM model ($73.14 \pm 1.10$ km/s/Mpc), and another  which only assumes the FLRW metric ($74.56 \pm 1.61$ km/s/Mpc). The study also stresses the importance of the supernova magnitude calibration ($M_B$) in cosmological inference and highlights the tension in $M_B$ when supernovae are calibrated either by CMB and BAO observations or the first two rungs of the cosmic distance ladder. This discrepancy, independent of the physics involved, suggests that models solely changing the Hubble flow and maintaining a sound horizon distance consistent with CMB, fail to explain the discrepancy between early- and late-time measurements of $H_0$.
}

\section{Introduction}
\label{sec:intro}

In the study of cosmology, one of the key quantities of interest is the Hubble constant, denoted as $H_0$. It represents the current rate of expansion of the universe and plays a crucial role in determining its age and evolution. Over the years, various methods have been employed to measure $H_0$, including observations of the cosmic microwave background (CMB), baryon acoustic oscillations (BAO), and the use of astronomical distance indicators such as Type Ia supernovae and Cepheids variables.

In recent years, the tension between different measurements of $H_0$ has garnered significant attention. While the Planck satellite mission \cite{Aghanim:2018eyx}, which analyzed the CMB, reported a value of $H_0$ around 67 km/s/Mpc, independent measurements using the local distance ladder technique, such as the SH0ES collaboration \cite{Riess:2021jrx}, have consistently obtained higher values, closer to 73 km/s/Mpc. This discrepancy, significant at the 5$\sigma$ level and often referred to as the ``Hubble tension,'' has led to considerable interest in exploring cosmological models that can reconcile these conflicting results.

Here, we aim to understand the implications of the local calibration of the luminosity of Type Ia supernovae. After reviewing the basic background in Section~\ref{sec:cg}, in Section~\ref{sec:local} we will start with the 46-dimensional analysis by SH0ES and extend it to an arbitrary cosmographic model. This will allow us to obtain a determination of local $H_0$ that only assumes the validity of the standard flat $\Lambda$CDM model, and another one that only assumes the FLRW metric, that is, large-scale homogeneity and isotropy.
We proceed in Section~\ref{sec:cosmoMB} to highlight the importance of considering the calibration of the supernova magnitude $M_B$ during cosmological inference, emphasizing the necessity to test the consistency of a given cosmological model with respect to $M_B$.
Finally, in Section~\ref{sec:tension} we discuss the $``M_B$ tension,'' the fact that the supernova luminosity either gets calibrated by CMB and BAO observations or by the first two rungs of the cosmic distance ladder, providing inconsistent constraints.
Our conclusions are presented in Section~\ref{sec:conclu}.

\section{Curved $\boldsymbol{w}$CDM cosmography}
\label{sec:cg}

Given the luminosity distance $d_L$, the apparent magnitude is given by:
\begin{equation} \label{mB}
m_B(z) = 5\log_{10}\left[\frac{d_L(z)}{1 \text{Mpc}} \right] +25 + M_B \,, 
\end{equation}
and the distance modulus by:
\begin{align} \label{mu}
\mu(z)= m_B(z) - M_B \,,
\end{align}
where $M_B$ is the absolute magnitude of Type Ia supernovae.
The luminosity distance is a prediction of a cosmological model and, assuming the FLRW metric, is:
\begin{align} \label{dL}
d_L(z) = (1+z) \frac{c}{ H_0} \Omega_{k0}^{-1/2} \sinh\left [ \Omega_{k0}^{1/2} \int_0^z \frac{\d \bar z}{E(\bar z)}  \right ] \,,
\end{align}
where $H=\dot a(t)/a(t)$ describes the expansion history of the universe and $a(t)$ is the scale factor. For a curved $w$CDM model it is:
\begin{align}
\frac{H^2(z)}{H^2_0} = E^2(z)   = \Omega_{\rm m0} (1+z)^3 + \Omega_{\rm de 0}(1+z)^{3(1+w)} + \Omega_{k 0} (1+z)^2 \,,
\end{align}
where we neglected radiation as it is inconsequential at redshifts lower than 100.
In the previous equation the subscript 0 denotes the present-day value of the corresponding quantity, $\Omega_{\rm m0}$ is the density parameter for cold dark matter (zero pressure), $\Omega_{\rm de0}$ the one for a dark energy fluid with equation of state parameter $w$, and $\Omega_{k0}=-k c^2/H_0^2$ represents the spatial-curvature contribution to the Friedmann equation ($a(t_0)=1$). It is $\Omega_{\rm m0}+\Omega_{\rm de0}+\Omega_{k0}=1$.

Alternatively, using a cosmographic approach one has \cite{Xia:2011iv}:
\begin{align} \label{dLcg}
d_L(z) &=  \frac{c z}{H_0} \left [ 1 + \frac{1-q_0}{2}\, z  -\frac{1- q_0-3 q_0^2+j_0 -\Omega_{k0} }{6} \, z^2  + O(z^3) \right] \,,
\end{align}
where the Hubble constant $H_0$, the deceleration parameter $q_0$ and the jerk parameter $j_0$ are defined, respectively, according to:
\begin{align}
H_0=\left. \frac{\dot a(t)}{a(t)} \right|_{t_0} \,,
\qquad
q_0=\left. \frac{- \ddot a(t)}{H^2(t) a(t)} \right|_{t_0}  \,,
\qquad
j_0=\left.\frac{\dddot a(t)}{H^3(t) a(t)}\right|_{t_0} \,.
\end{align}
Cosmography is a model-independent approach in the sense that it does not assume a specific model as it is based on the Taylor expansion of the scale factor. However, this does not mean that its parameters do not contain cosmological information.
For example, in the case of curved $w$CDM cosmologies, $q_0$ and $j_0$ are given by:
\begin{align}
q_0 &
= \frac{\Omega_{\rm m 0}}{2} + \frac{1+3 w}{2} \Omega_{\rm de 0}
\stackrel{\text{flat}}{=} \frac{1}{2}+ \frac{3}{2} w(1-\Omega_{\rm m 0})
\stackrel{\Lambda}{=}   \frac{3}{2} \Omega_{\rm m 0} -1
\stackrel{\text{fid}}{=} -0.55 \,,  \label{q0}  \\
j_0 & = \Omega_{\rm de 0}+ \Omega_{\rm m 0}  + \frac{9}{2}w(1+w) \Omega_{\rm de 0}
\stackrel{\text{flat}}{=}1 + \frac{9}{2}w(1+w) (1-\Omega_{\rm m 0}) 
\stackrel{\Lambda}{=} 1 \,,  \label{j0}
\end{align}
where in the last equality of \eqref{q0} the concordance value of $\Omega_{\rm m 0}=0.3$ was adopted.

Finally, when determining the luminosity distance, one should adopt both the peculiar velocity-corrected CMB-frame redshift $z$ (also called Hubble diagram redshift) and the heliocentric redshift $z_{\rm hel}$ so that the luminosity distance is given by~\cite{Calcino:2016jpu}:
\begin{align}
d_L(z, z_{\rm hel})= \frac{1+z_{\rm hel}}{1+z}d_L(z) \,.
\end{align}

\section{Local determination of $H_0$}
\label{sec:local}

The SH0ES collaboration measures the Hubble constant in the redshift range $0.023<z<0.15$, where the minimum redshift is large enough in order to reduce the impact of cosmic variance by the local large-scale structure \cite{Marra:2013rba,Camarena:2018nbr} and the maximum redshift is small enough to provide a late-time local measurement of the Hubble constant and reduce the impact of cosmology in its determination. Ideally, one would want a cosmology-independent local constraint.
SH0ES traditionally fits the value of $H_0$ in \eqref{dLcg}, but fixes the values of the higher-order parameters to their fiducial values of $q^{\rm fid}_0=-0.55$ and $j^{\rm fid}_0=1$, using information from observations beyond the local universe, like higher-$z$ supernovae.

In \cite{Camarena:2019moy,Camarena:2021jlr} we obtained a local determination of $H_0$ without fixing the deceleration parameter $q_0$.
Here, we will improve upon our previous results by extending the analysis of \cite{Riess:2021jrx} using their latest data.\footnote{\url{https://github.com/PantheonPlusSH0ES/DataRelease}}
Specifically, we adopt the $\chi^2$ statistic:
\begin{align} \label{chi2}
\chi^2(\theta, q_0, j_0) = \big (y(q_0, j_0)-L \theta \big )^{\rm T} C_y^{-1} \big (y(q_0, j_0)-L \theta \big ) \,,
\end{align}
where $y$ is the 3492-dimensional data vector, $\theta$ is the 46-d parameter vector considered by SH0ES, $L$ is the 46$\times$3492 equation matrix, so that the model is $L\theta$, and $C_y$ is the 3492$\times$3492 covariance matrix.
The last elements of $y$ are about the 277 Hubble flow supernovae (nhf) that are used to determine $H_0$:
\begin{align}
y^{\textrm{nhf}}_i &= m_{B, i} - 5 \log_{10} H_0 d_L - 25  \\
&=    m_{B, i} - 5 \log_{10} c z_i \left [ 1 + \frac{1-q_0}{2}\, z_i  -\frac{1- q_0-3 q_0^2+j_0 -\Omega_{k0} }{6} \, z_i^2  \right] -25 \nonumber \,,
\end{align}
and in Eq.~\eqref{chi2} we let the components $y^{\textrm{nhf}}_i$ be functions of $q_0$ and $j_0$.\footnote{We set $\Omega_{k0}=0$, but the latter is degenerated with $j_0$, see discussion below.}

\begin{figure}[t]
\centering 
\includegraphics[width= \textwidth]{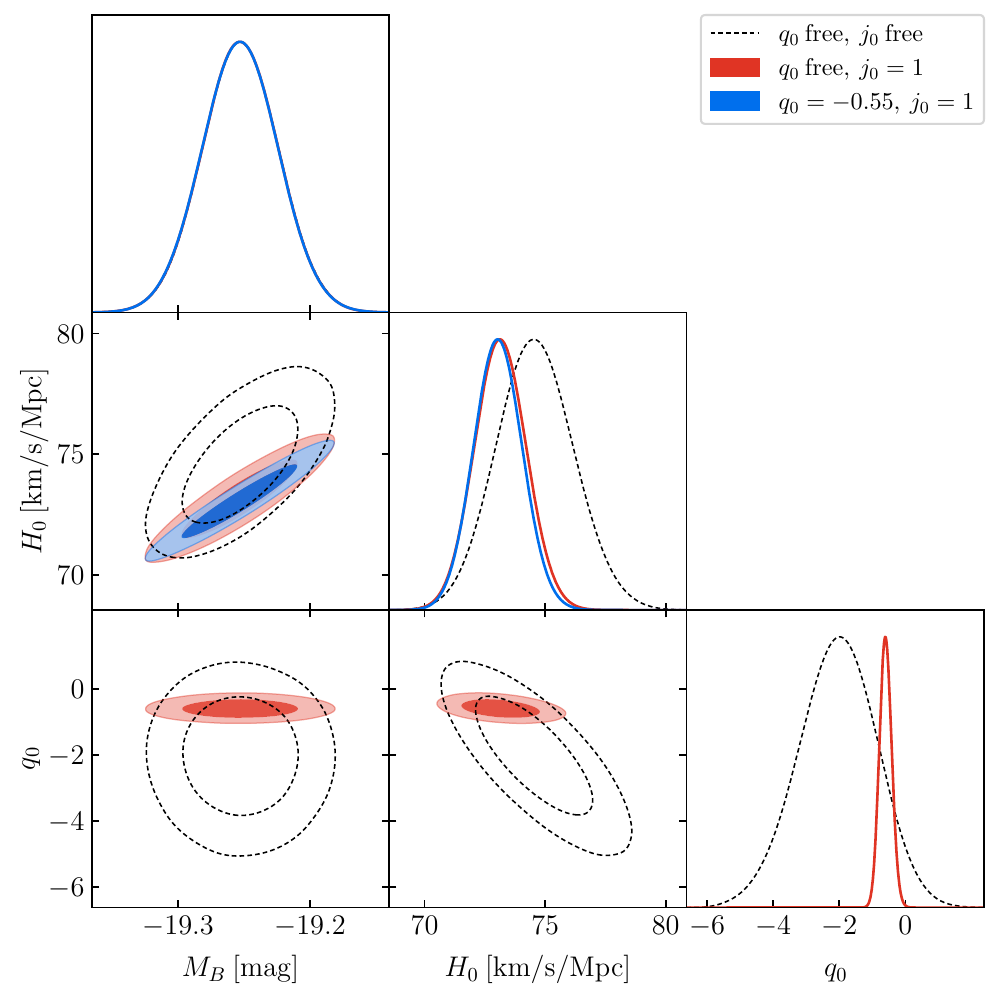}
\caption{Marginalized 68.3\% and 95.4\% credible level constraints on the supernova Ia absolute magnitude $M_B$, the Hubble constant $H_0$ and the deceleration parameter $q_0$ from the 48-dimensional analysis of Eq.~\eqref{chi2ma} using the latest data from the SH0ES collaboration, that is, the first two rungs of the distance ladder together with the 277 Hubble flow supernovae in $0.023<z<0.15$.
Blue shows the SH0ES baseline analysis, red the analysis that only assumes the flat $\Lambda$CDM model and promote $q_0$ (or, equivalently, $\Omega_{\rm m 0}$) to a free parameter, and the empty contours show the analysis that only assumes the FLRW metric, that is, large-scale homogeneity and isotropy (both $q_0$ and $j_0$ are free, the latter is marginalized over).}
\label{fig:H0q0}
\end{figure}

The baseline analysis by SH0ES adopts $\chi^2(\theta, q^{\rm fid}_0, j^{\rm fid}_0)$, so that $y$ is constant and the model is linear.
In this case the best-fit model and its covariance matrix are:
\begin{align}
\theta_{\rm bf}(q_0, j_0) &= \big ( L^{\rm T} C_y^{-1} L \big )^{-1}L^{\rm T} C_y^{-1} \, y(q_0, j_0) \,, \\
C_\theta &= \big ( L^{\rm T} C_y^{-1} L \big )^{-1} \,,
\end{align}
from which we can note that the best fits depend on $q_0$ and $j_0$, but their covariance matrix does not.
Using the properties of multi-variate Gaussian distributions (the model is linear), it is then straightforward to plot the marginalized constraints on any combination of parameters by considering the eigenvectors and eigenvalues of the covariance matrix.
This is equivalent to performing a Bayesian analysis with wide flat priors, as done in \cite{Riess:2021jrx}.
In Fig.~\ref{fig:H0q0} we show the marginalized constraints on $M_B$ and $H_0$.

If we now do not fix $q_0$ and $j_0$, the model used in Eq.~\eqref{chi2} becomes nonlinear and a full Bayesian analysis is necessary. One possible approach is to approximate the posterior as a Gaussian distribution via its Fisher approximation, $F_{\alpha, ij} = \frac{1}{2} \frac{\partial^2 \chi^2}{\partial \alpha_i \partial \alpha_j}$,
where $\alpha= \{ \theta, q_0, j_0 \} $ is the 48-d parameter vector, and $C_\alpha=F_\alpha^{-1}$ is the covariance matrix on the parameters (the inverse of the Fisher matrix).\footnote{Note that $\frac{1}{2} \frac{\partial^2 \chi^2}{\partial \theta_i \partial \theta_j}$ exactly gives $C^{-1}_\theta$ as, in this case, the model is linear.}
However, the posterior will likely show a non-Gaussian character in the $j_0$ direction as the latter is expected to be poorly constrained by supernovae at $z<0.15$. Consequently one should perform a computationally demanding 48-d Bayesian analysis.

Fortunately, we can  exploit the fact that the model is linear in all parameters but $q_0$ and $j_0$ to substantially speed up the estimation of the posterior.
We can indeed marginalize analytically on all the parameters we are not interested in:
\begin{align}
&\chi^2_{\rm marg} (M_B, H_0, q_0, j_0) = \chi^2 \big (\theta_{\rm bf}(q_0, j_0), q_0, j_0 \big ) +  V C_{M_B, H_0}^{-1} V \,, \label{chi2ma} \\
&V = \{ M_B, H_0 \} -\{M_B^{\rm bf}(q_0, j_0) , H_0^{\rm bf}(q_0, j_0) \}   \,,
\end{align}
where $M_B^{\rm bf}$ and $H_0^{\rm bf}$ comes from the corresponding entries of $\theta_{\rm bf}(q_0, j_0)$ \cite{Riess:2021jrx}, and $C_{M_B, H_0}$ is obtained by removing all the rows and columns of $C_\theta$ expect the ones relative to $M_B$ and $H_0$, as this operation is equivalent to marginalizing the posterior over all the other linear parameters.

In Figure~\ref{fig:H0q0} and Table~\ref{tab:H0q0} we compare the baseline results by SH0ES with our results.
It is interesting to note that the analysis relative to $q_0$ free and $j_0=1$ provides a very competitive constraint that is independent of the particular $\Lambda$CDM model as it assumes just flat FLRW metric and  cosmological constant.
It features a 1.5\% uncertainty, as compared with the 1.4\% uncertainty of the baseline analysis.
In particular, marginalizing over $q_0$ is equivalent to marginalizing over $\Omega_{\rm m 0}$.
Indeed, their relation is linear, so that the same wide flat prior is adopted, and the relative difference of $\Lambda$CDM  with respect to the corresponding cosmographic model, see Eq.~\eqref{q0}, is $\approx 10^{-6}$ at $z=0.15$.

\begin{table}[t]
\centering
\setlength{\tabcolsep}{10pt}
\renewcommand{\arraystretch}{1.3}
\begin{tabular}{l|c|c|c}
Case & $M_B$ (mag) & $H_0$ (km/s/Mpc) & $q_0$  \\
\hline
 $q_0=-0.55$, $j_0=1$ & $-19.253 \pm 0.029$ & $73.04 \pm 1.04$ (1.4\%) & --  \\
 $q_0$ free, $j_0=1$ & $-19.253 \pm 0.029$ & $73.14 \pm 1.10$  (1.5\%) & $-0.61 \pm 0.18$ \\
 $q_0$ and $j_0$ free & $-19.252 \pm 0.029$ & $74.56 \pm 1.61$  (2.2\%) & $-2.0 \pm 1.2$ \\
\end{tabular}
\caption{Marginalized constraints  (mode and 68\% uncertainty) for the three  cosmographic analyses of Fig.~\ref{fig:H0q0}.
The first line corresponds to the baseline analysis by SH0ES \cite{Riess:2021jrx}, providing the same constraints. The second analysis gives a constraint that is marginalized over any flat $\Lambda$CDM model.
The last analysis only assumes the FLRW metric. Following \cite{Riess:2021jrx}, an additional systematic error of 0.3 km/s/Mpc was added to the uncertainty of $H_0$.}
\label{tab:H0q0}
\end{table}

Next, by adopting an improper flat prior for $j_0$, one obtains a measurement that only assumes the FLRW metric, that is, large-scale homogeneity and isotropy. It is interesting to note that in Eq.~\eqref{dLcg} the degenerate combination $j_0-\Omega_{k0}$ appears so that by marginalizing over $j_0$, we are effectively marginalizing also over spatial curvature. The data poorly constrain this combination, giving $j_0 - \Omega_{k0}= 28^{+39}_{-28}$.

Finally, we found that the deceleration parameter is in good agreement  with the value relative to the standard model, contrary to what we found in \cite{Camarena:2019moy,Camarena:2021jlr} with the Pantheon and Supercal datasets. The difference is likely due to the improved modeling of uncertainty and redshift corrections in the Pantheon+ dataset~\cite{Brout:2022vxf}.

\section{Cosmological inference with the local prior on $M_B$}
\label{sec:cosmoMB}

The latest combined data release by the SH0ES and Pantheon+ collaborations significantly improved the robustness and consistency of cosmological analyses that include a local determination of the Hubble constant.
Now, the distance calibration from 37 Cepheids is provided for the 42 supernovae in the Cepheid host galaxies,\footnote{Some galaxies hosted more than one supernova. Note also that some supernovae have been observed by more than one telescope so that in Pantheon+ there are actually 47 supernova entries that are calibrated by Cepheid distances.}
together with the combined supernova and Cepheid covariance matrix~\cite{Brout:2022vxf}.
This provides an absolute calibration for the supernova Ia absolute magnitude $M_B$ as the likelihood is now given by:
\begin{align}
V_i&= \begin{cases} 
m_{B,i}-M_B -  \mu(z_i) & \text{if Hubble diagram supernova,} \\
m_{B,i}-M_B -  \mu_{{\rm ceph},i} & \text{if supernova in Cepheid host,}  
\end{cases}  \nonumber\\
\chi_\text{sne}^{2} &=  V^{\rm T} \big (  \Sigma_\text{sne}+\Sigma_\text{ceph}  \big )^{-1} V \,,\label{chi2sne}
\end{align}
where $\Sigma_\text{ceph}$ has nonzero entries only for the supernovae in Cepheid hosts.
The ``Hubble diagram'' (HD) supernovae are the 1466 supernovae -- corresponding to 1580 measurements -- that are used to constrain cosmological models but not to calibrate $M_B$. These have a redshift $z>0.01$, again to remove the cosmic variance from the local structure.
The 42 supernovae (47 light curves) in Cepheid host galaxies are instead used to calibrate $M_B$, whose calibration is then propagated to the HD supernovae. This breaks the degeneracy between $M_B$ and the $5 \log_{10} H_0$ term inside the distance modulus $\mu(z)$, see Eq.~\eqref{mu}.

The above methodology is ideal as it includes the correlations between Cepheid distances and calibrating and HD supernovae, ensuring an accurate representativeness of the cosmic distance ladder in  cosmological inference~\cite{Camarena:2019moy, Benevento:2020fev, Camarena:2021jlr, Efstathiou:2021ocp,Greene:2021shv}. Indeed, if one just uses HD supernovae, together with a local prior on $H_0$, one will i) double count the 277 low-redshift supernova measurements that were used to obtain the prior on $H_0$, ii) assume the validity of cosmography, in particular fixing the deceleration parameter to the standard model value of $q_0=-0.55$, iii) not include in the analysis the fact that $M_B$ is constrained by the local calibration, an information which would otherwise be neglected in the analysis, biasing both model selection and parameter constraints~\cite{Camarena:2021jlr}.

The approach of Eq.~\eqref{chi2sne} allows to carry out a variety of analyses. For example, one can only consider the 238 HD supernovae (but 277 supernova measurements) that are used to determine the Hubble constant and determine $H_0$ according to various fiducial models, that is, varying $q_0$ and $j_0$ in the case of cosmography or varying $\Omega_{\rm m 0}$ and $w$ in the case of $w$CDM models.
This approach is equivalent to the one of the previous Section, which only uses the quantities $L$, $y$ and $C$ from SH0ES. Here, we preferred to adopt the methodology of the previous Section because we noticed that the Pantheon+ dataset produces a value of $H_0$ which is $0.5$~km/s/Mpc higher than the SH0ES baseline value when considering only the 277 low-redshift supernova measurements that are used to determine the Hubble constant.\footnote{We contacted the authors of \cite{Riess:2021jrx} and \cite{Brout:2022vxf}; according to Adam Riess there may be an inconsistency between the production and application of the covariance matrix of Cepheid/SN Ia host distances in the Pantheon+ fitting code. This issue is under investigation.}

The most interesting use of Eq.~\eqref{chi2sne} is the possibility of constraining the Hubble constant conditionally to a particular cosmological model, bypassing completely the use of cosmography. In this case the tension in the high- and low-redshift constraint on $H_0$ is traduced into a tension between the local calibration of $M_B$ and the one induced by the CMB via BAO observations.
This highlights the important of using the new Pantheon+ \& SH0ES likelihood.

\begin{figure}[t]
\centering 
\includegraphics[width= 0.48 \textwidth]{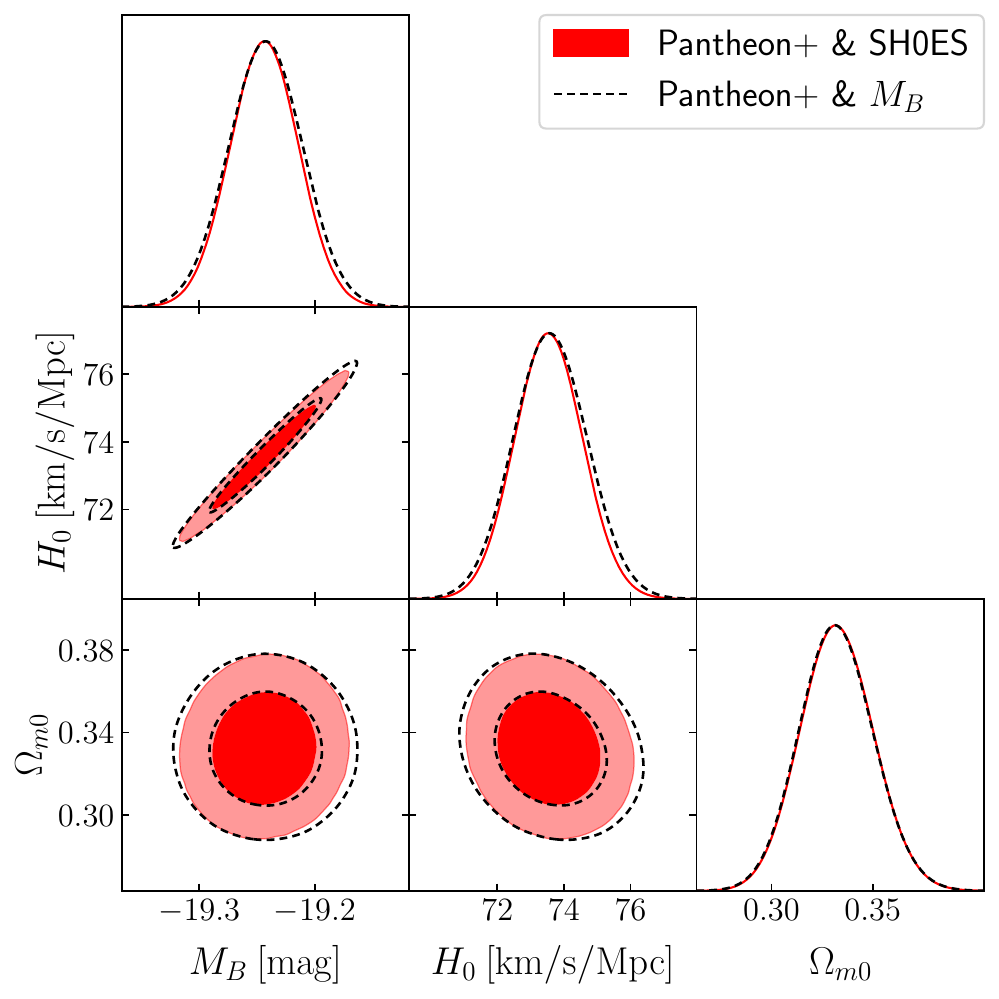}
\includegraphics[width= 0.48 \textwidth]{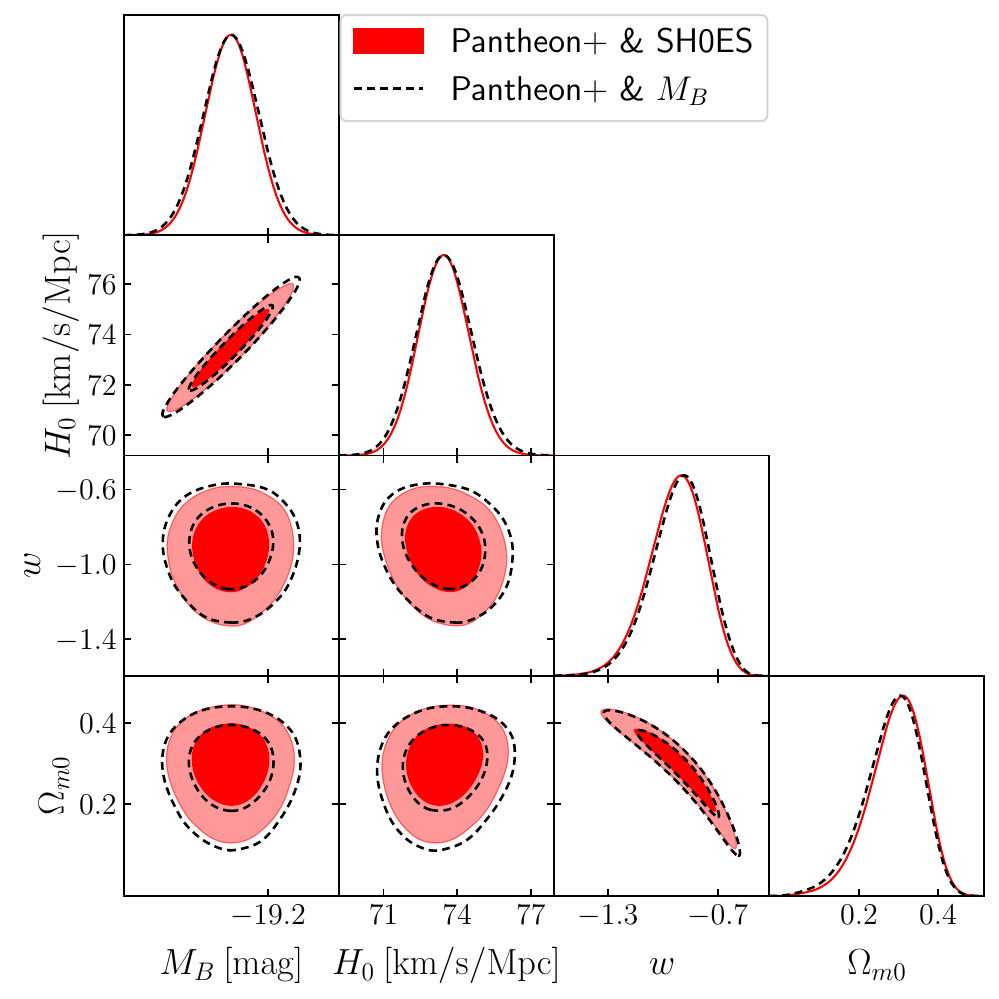}
\caption{Comparison between the analysis that uses the Pantheon+ \& SH0ES likelihood and the one that uses the Pantheon+ likelihood with the Gaussian prior on $M_B$ from Eq.~\eqref{MBprior1}, for the flat $\Lambda$CDM model (left) and the flat $w$CDM model (right).}
\label{fig:Mprior}
\end{figure}

In order to offer an easy way to implement the absolute calibration of $M_B$, here we apply the demarginalization method introduced in \cite{Camarena:2019moy}  to obtain an effective prior on $M_B$.
As target for the demarginalization procedure, we consider both the constraint $H_0 = 73.6 \pm 1.1$ km/s/Mpc for the flat $\Lambda$CDM model from \cite{Brout:2022vxf} and a value $0.5$~km/s/Mpc lower, because of the discrepancy mentioned above when using the Pantheon+ \& SH0ES likelihood.\footnote{This approach was suggested by Adam Riess.}
We use the 1580 HD supernova entries and obtain:
\begin{align} 
M_B &= -19.243 \pm 0.032 \text{ mag \quad for \quad} H_0 = 73.6 \pm 1.1 \text{km/s/Mpc} \,, \label{MBprior1} \\
M_B &= -19.258 \pm 0.032 \text{ mag \quad for \quad} H_0 = 73.1 \pm 1.1 \text{km/s/Mpc} \,. \label{MBprior2}
\end{align}

As a cross-check, using Eq.~\eqref{MBprior1} together with the Pantheon+ likelihood, we obtain a good agreement with the results from \cite{Brout:2022vxf}:\footnote{Ref.~\cite{Brout:2022vxf} adopted a prior $\Omega_{\rm m 0}>0.1$, affecting their results in that parameter space region.}
\begin{align}
H_0 &= 73.6 \pm 1.1   \qquad \Omega_{\rm m 0} =0.333 \pm 0.018    \qquad &\text{($\Lambda$CDM)} \,, \nonumber \\
H_0 &=73.5 \pm 1.1    \qquad \Omega_{\rm m 0} =0.306^{+0.063}_{-0.076}    \qquad  w=-0.88\pm0.15 \qquad &\text{($w$CDM)} \,. \nonumber
\end{align}
Fig.~\ref{fig:Mprior} shows a full comparison between the analysis that uses the Pantheon+ \& SH0ES likelihood and the one that uses the Pantheon+ likelihood with the Gaussian prior on $M_B$. Constraints and correlations are reproduced accurately.
Furthermore, the use of the Gaussian prior has the advantage that one can marginalize analytically over $M_B$, see \cite{Camarena:2021jlr} for details.

\section{$M_B$ tension between SH0ES and Planck}
\label{sec:tension}

Several theoretical models have been proposed to resolve the discrepancy between the CMB determination of the Hubble constant by Planck~\cite{Aghanim:2018eyx} and the local measurement by SH0ES \cite{Riess:2021jrx}. These models modify the $\Lambda$CDM cosmology either in the late or early universe. Early-time modifications of the standard cosmological model typically accommodate a larger $H_0$ value by reducing the sound horizon scale $r_s$ during recombination, whereas late-time modifications typically increase the Hubble rate at low redshifts \cite{Knox:2019rjx,Schoneberg:2021qvd,Shah:2021onj}.
While no compelling evidence for physics beyond the $\Lambda$CDM model has been found, late-time modifications of the standard paradigm face an additional problem: the $M_B$ tension.

The local determination of the Hubble constant provided by the SH0ES collaboration relies on a set of astrophysical distances that are used to calibrate the absolute magnitude of supernovae. In \cite{Camarena:2019rmj} we built  a model-independent version of the inverse distance ladder, which uses BAO distances and a prior on $r_s$, coming from the Planck 2018 data release~\cite{Aghanim:2018eyx}, to effectively calibrate the supernovae. Our analysis demonstrated that the absolute magnitude obtained from the
cosmic distance ladder~\cite{Camarena:2019moy} is in tension with the underlying calibration on $M_B$ provided by the combination of CMB + BAO, see Tab.~\ref{tab:LCDM_Pantheon}. Since the latter constitutes a standard ruler that relies on the value adopted for $r_s$, the discrepancy in $M_B$ suggests that, independent of the physics, models that solely change the Hubble flow, while keeping a sound horizon distance consistent with the one inferred from CMB under the assumption of a $\Lambda$CDM cosmology, will fail at explaining the discrepancy between the early and late times.

\begin{table}[t]
\centering
\setlength{\tabcolsep}{7pt}
\renewcommand{\arraystretch}{1.3}
\begin{tabular}{l|c|c|c}
\hline
\hline
 $\Lambda$CDM analysis & $M_B$ (mag) & $H_0$ (km/s/Mpc) & $\Omega_{m0}$  \\
\hline
 Pantheon+ $\&$ SH0ES 22 & $-19.244 \pm 0.029$ & $73.55 \pm 1.02$ $ (\pm 0.3$) & $0.332 \pm 0.018$   \\
 Pantheon+ $\&$ CMB + BAO & $-19.438 \pm 0.007$ & $67.39 \pm 0.19$   & $0.315 \pm 0.002$ \\
  Tension  & $6.5\sigma$ & $5.7\sigma$  & --\\
  \hline
\hline
Model-independent analysis & $M_B$ (mag) & $H_0$ (km/s/Mpc) &  $q_0$ \\
\hline
 Pantheon $\&$ SH0ES 21 \cite{Camarena:2021jlr} & $-19.244 \pm 0.037$ & $74.30 \pm 1.45$ & $-0.91 \pm 0.22$   \\
 Pantheon $\&$ CMB + BAO \cite{Camarena:2019rmj} & $-19.401 \pm 0.027$ & $69.71 \pm 1.28$   &  $-1.09 \pm 0.29$ \\
  Tension  & $3.4\sigma$ & $2.4\sigma$  & --  \\
\hline
\hline
\end{tabular}
\caption{The $M_B$ tension.
Top: marginalized constraints adopting the $\Lambda$CDM model. The first line corresponds to the results by \cite{Brout:2022vxf}, providing the same constraints.
The second line shows the constraints that are obtained when including the full CMB likelihood together with the latest BAO measurements instead of the local prior on Cepheid host galaxies by SH0ES.
Bottom: as above but for the Pantheon dataset \cite{Scolnic:2017caz} and adopting the model-independent (inverse) distance ladder of \cite{Camarena:2019rmj}.}
\label{tab:LCDM_Pantheon}
\end{table}

Indeed, the inability of late-time models to adequately explain the Hubble tension has been clearly demonstrated by several analyses \cite{Benevento:2020fev,Efstathiou:2021ocp}, including the analysis presented in \cite{Camarena:2021jlr}. This particular analysis employs a toy model, referred to as the `hockey-stick' model, wherein the dark energy component rapidly undergoes a phantom transitions at low redshift. This illustrates that the $M_B$ tension must be addressed in order to resolve the Hubble discrepancy effectively.
As shown in~\cite{Camarena:2021jlr}, when analyzed with a prior on $H_0$ together with CMB and BAO data, the hockey-stick model, $hs$CDM, provides a value of the Hubble constant consistent with the local observation. Nevertheless, the underlying calibration of $M_B$ that is obtained in the analysis still is in tension with the calibration inferred from the cosmic distance ladder, showing that $hs$CDM does not really restore the agreement between early and late times distances. This issue, and spurious conclusions, can be avoided if instead a prior on $M_B$ is assumed.\footnote{Alternatively, one should build a likelihood containing all the rungs of the cosmic distance ladder~\cite{Greene:2021shv}. For a more detailed discussion, see Sec.~\ref{sec:cosmoMB}.} Indeed, the analysis of the $hs$CDM model under the assumption of this prior explicitly shows the failure of the model to alleviate the $H_0$ ($M_B$) tension.
Similar conclusions hold for void models, which try to explain the local higher expansion rate by placing the observer inside a large void. As far as the observed luminosity distance-redshift relation is concerned, these models are very similar to the $hs$CDM model and are ruled out for the very same reasons~\cite{Camarena:2022iae}.

A category of models that potentially resolve the $M_B$ tension encompasses those that feature a transition in supernova brightness at the extremely low redshifts of the first two rungs -- specifically, before entering the Hubble flow with supernovae in the range $0.023<z<0.15$.
This transition could be realized by a sudden shift in the effective gravitational constant by approximately 10\% around 50–150 million years ago, consequently altering the Chandrasekhar mass and thereby the supernova luminosity \cite{Marra:2021fvf,Alestas:2021luu}. While this proposal conveniently resolves the Hubble tension, it also yields a series of testable predictions that could either support or disprove it \cite{Perivolaropoulos:2022vql}.
Physical mechanisms that might trigger such an ultra-late gravitational transition include a first-order scalar-tensor theory phase transition from an early false vacuum, corresponding to the measured value of the cosmological constant, to a new vacuum with lower or zero vacuum energy~\cite{Abdalla:2022yfr}.

\begin{figure}[t]
\centering 
\includegraphics[width= 0.8\textwidth]{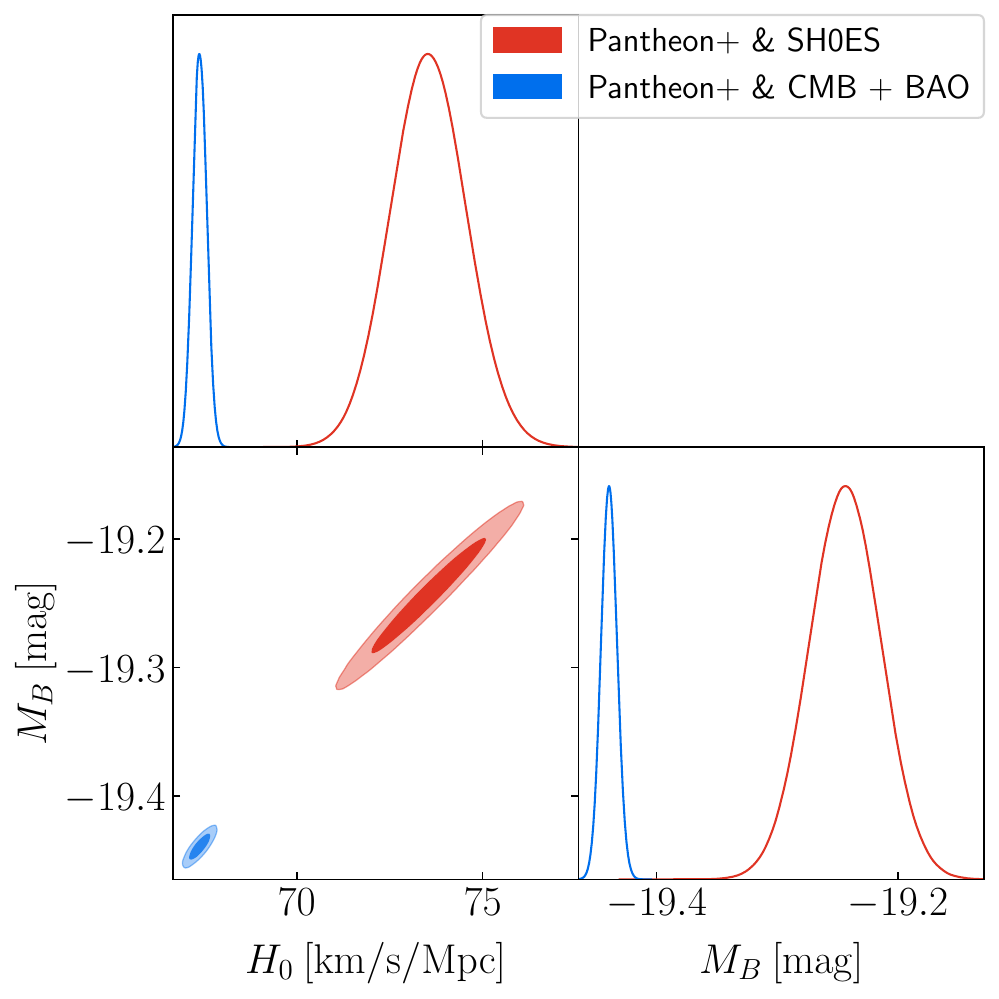}
\caption{
Marginalized constraints from the cosmological equivalents to the cosmic distance ladder (Pantheon+ \& SH0ES data) and the inverse distance ladder (Pantheon+ in combination with BAO and CMB) when assuming the $\Lambda$CDM model as fiducial model of the universe. This figures shows how the tension on the Hubble constant can be interpreted as a tension in the calibration of the supernova brightness.
}
\label{fig:LCDM_Pantheon}
\end{figure}

We update the current significance of the $M_B$ tension using the latest data and adopting the fiducial $\Lambda$CDM model. Using BAO data from BOSS DR12 \cite{BOSS:2016wmc}, eBOSS ELG \cite{deMattia:2020fkb}, eBOSS LRG \cite{Bautista:2020ahg,Gil-Marin:2020bct}, SDSS MGS \cite{Ross:2014qpa}, and WiggleZ \cite{Kazin:2014qga},  high-$\ell$ TT+TE+EE, low-$\ell$ TT, and low-$\ell$ EE CMB data from Planck 2018~\cite{Aghanim:2018eyx},\footnote{For simplicity, we used the compressed version of high-$\ell$ CMB data.} and the latest Pantheon+ \& SH0ES release, we constrain the $\Lambda\mathrm{CDM}$ model considering the cosmological equivalents to the cosmic distance ladder (Pantheon+ \& SH0ES data) and the inverse distance ladder (Pantheon+ in combination with BAO and CMB). The results of this analysis are presented in Fig.~\ref{fig:LCDM_Pantheon} and Tab.~\ref{tab:LCDM_Pantheon}. We used CosmoSIS and CAMB \cite{Lewis:1999bs,Howlett:2012mh} to carry out the analyses.
We find that the tension on $M_B$ exceeds the 6$\sigma$ level.

\section{Conclusions}
\label{sec:conclu}

In this study, we have examined three distinct yet interconnected topics.
Firstly, we extended the analysis conducted by the SH0ES collaboration to determine the local value of the Hubble constant. We broadened their 46-dimensional analysis to cosmography with arbitrary values of the deceleration and jerk parameters.
\begin{svgraybox}
We found that the latest Pantheon+ \& SH0ES data provides a competitive constraint that solely assumes the validity of the standard flat $\Lambda$CDM model:
$$
73.14 \pm 1.10  \; (1.5\%) \text{ km/s/Mpc},
$$
and a model-independent constraint that only assumes the FLRW metric:
$$
74.56 \pm 1.61 \;  (2.2\%) \text{ km/s/Mpc}.
$$
\end{svgraybox}

Next, we underscored the importance of considering the calibration of the supernova magnitude $M_B$ by the first two rungs of the distance ladder when conducting cosmological inference. Indeed, $M_B$ is a crucial astrophysical parameter and should not be integrated out, but examined to understand the consistency of a given cosmological model.

This discussion  led us to the issue of the $``M_B$ tension.''
The supernova luminosity either gets calibrated by CMB and BAO observations or by the first two rungs of the cosmic distance ladder.
\begin{svgraybox}
We found that, assuming the standard flat $\Lambda$CDM model, the two constraints on $M_B$ are in tension at the 6.5$\sigma$ level.
\end{svgraybox}
The discrepancy in $M_B$ suggests that, independent of the physics involved, models that solely change the Hubble flow, while maintaining a sound horizon distance consistent with the one inferred from CMB, fail to explain the discrepancy between the early- and late-time measurements of $H_0$.

\begin{acknowledgement}
We are grateful to Adam Riess for his valuable feedback regarding the SH0ES data release.
DC thanks the Robert E.~Young Origins of the Universe Chair fund for its generous support. VM thanks CNPq (Brazil, 307969/2022-3) and FAPES (Brazil, TO 365/2022, 712/2022, 976/2022, 1020/2022, 1081/2022) for partial financial support.
The authors acknowledge the use of the computational resources of the Sci-Com Lab of the Department of Physics--UFES.
\end{acknowledgement}

\bibliographystyle{utcaps}
\bibliography{biblio}

\end{document}